\documentclass[12pt,preprint]{aastex}

\shorttitle{The $\lambda$10830 \ion{He}{1} Line Among M13 Red Giants}
\shortauthors{Smith, Dupree, \& Strader}

\begin{document}

\title{A Study of the $\lambda$10830 \ion{He}{1} Line Among Red Giants in 
Messier 13\footnote{The data presented herein were obtained at the W.M. Keck 
Observatory, which is operated as a scientific partnership among the California
Institute of Technology, the University of California and the National 
Aeronautics and Space Administration. The Observatory was made possible by the 
generous financial support of the W.M. Keck Foundation.}} 

\author{Graeme H. Smith}
\affil{University of California Observatories, Lick Observatory, Department 
 of Astronomy \& Astrophysics, UC Santa Cruz, 1156 High St., Santa Cruz, CA 
 95064, USA}
\email{graeme@ucolick.org}

\author{Andrea K. Dupree}
\affil{Harvard-Smithsonian Center for Astrophysics, Cambridge, MA 02138, USA}
\email{dupree@cfa.harvard.edu}

\author{Jay Strader}
\affil{Department of Physics and Astronomy, Michigan State University, 
 East Lansing, MI 48824, USA}
\email{strader@pa.msu.edu}

\begin{abstract}
Two properties of Messier 13 are pertinent to the study of mass loss among 
metal-poor stars and the chemical evolution of globular clusters: (i) an 
extended blue horizontal branch, which seems to demand mass loss from red 
giant progenitor stars and possibly an enhanced helium abundance, and 
(ii) the presence of internal abundance inhomogeneities 
of elements in the mass range from C to Al. A popular explanation for this 
second phenomenon is that M13 was self-enriched by intermediate-mass asymptotic
giant branch (IM-AGB) stars of a type that may also have been able to instigate
helium enrichment. Spectra of the $\lambda$10830 absorption feature produced 
by \ion{He}{1} have been obtained by using the NIRSPEC spectrometer on the 
Keck 2 telescope for seven red giants in M13 chosen to have a range in 
$\lambda$3883 CN band strengths, oxygen and sodium abundances. Whereas these 
spectra do reveal the presence of fast winds among some M13 red giants, they 
provide little support for helium abundance differences of the type that 
might have been generated by a burst of IM-AGB star activity within the 
M13 protocluster.
\end{abstract}


\section{Introduction}

Globular clusters of the Milky Way are odd objects. The inhomogeneity of 
elements in the C-Al mass range within these systems indicates
that they have sustained heterogeneous self-enrichment, however elements such 
as iron and calcium are typically homogeneous among the stars within a given
cluster. The abundance inhomogeneities can be conveniently traced by the
strength of absorption bands of CN at 3883 \AA\ (e.g., Smith 1987). Many 
clusters, such as Messier 13, have a sub-population of CN-strong stars that 
are enhanced in Na and Al but are depleted in C and O relative to a CN-weak 
sub-population (e.g., Kraft et al. 1997; Sneden et al. 2004; Cohen \& 
Mel\'{e}ndez 2005; Johnson et al. 2005; Smith \& Briley 2006). 

Following the suggestion of Cottrell \& Da Costa (1981) quite a few efforts 
have been made to interpret the CN-strong stars (which tend to be enriched
in the elements N, Na and Al, but depleted in oxygen) as products of 
inhomogeneous enrichment of a globular cluster in processed ejecta from 
intermediate-mass asymptotic giant branch (IM-AGB) stars (e.g., Smith \& Norris
1982; Denissenkov et al. 1997; Ventura et al. 2001, 2002; Fenner et at. 2004; 
Bekki 2011), although other sources of element synthesis are arguably also 
feasible (e.g., Charbonnel 2009). Some clusters harboring chemical 
inhomogeneities may also show substructures in the color-magnitude diagram, 
such as multiple subgiant branches or main sequences, that have been 
interpreted as requiring inhomogeneities of He abundance (e.g., Norris 2004; 
Piotto et al. 2005; D'Antona et al. 2005; Bragaglia et al. 2010; di Ciscienzo 
et al. 2010). Within a self-enrichment scenario invoking IM-AGB stars a 
correlation between He and the CN sub-populations might be possible (e.g., 
D'Antona et al. 2002) if the enriching stars eject sufficient 
quantities of helium, the feasibility of which is the subject of some debate 
(e.g., Karakas et al. 2006; Catelan et al. 2009).

Helium abundances are hard to determine directly for globular clusters,
and are often inferred indirectly from the morphology of, or numbers of stars 
in, different regions of the color-magnitude diagram. Among blue horizontal 
stars where helium lines are observable in the spectrum, the derived abundance
can be subject to uncertainties of interpretation by diffusion effects, with 
the study of Villanova et al. (2009) being one exception.

In the near infrared there is a \ion{He}{1} triplet transition at 10830 \AA\ 
that can produce an absorption line from the upper chromosphere of late-type 
stars, both dwarfs and giants. It is a relatively weak feature that has to be 
observed at high resolution. Dupree et al. (1992) were the first to detect this
line in the spectrum of a metal-poor field giant: HD 6833 
(${\rm [Fe/H]} = -0.75$; Luck 1991). Previous observations of the \ion{He}{1} 
line among red giants by Zirin (1982) and O'Brien \&\ Lambert (1986) had been 
limited to Population~I giants. 

Smith et al. (2004) and Dupree et al. (2009) reported on the use of the NIRSPEC
spectrometer on the Keck 2 telescope to survey the $\lambda$10830 
\ion{He}{1} line among Population II stars in a variety of evolutionary states:
first ascent red giant branch (RGB) stars, red horizontal branch stars, 
asymptotic giant branch stars, and semi-regular variables. Two objectives of
these studies were to canvas the incidence of \ion{He}{1} absorption as a 
function of position in the color-magnitude diagram and to determine line 
profiles in a search for mass outflows. The spectra show a variety of 
phenomena. Whereas some giants do not show a \ion{He}{1} line at all, the 
majority do show some absorption. Among giants brighter than the horizontal 
branch, it is not uncommon for the \ion{He}{1} absorption profile to be 
extended on the blue side indicative of an outflow. Other stars show a more 
symmetrical \ion{He}{1} absorption line. Interestingly, evidence of outflows 
is detected among a sizable fraction of red horizontal branch stars.

Dupree et al. (1992) modelled the chromospheric \ion{He}{1} 10830 \AA\  
transition. Because of its large height of formation this line can reveal 
systematic bulk motions in the atmospheres of a luminous cool star. The 
\ion{He}{1} line is a pure chromospheric feature with no photospheric, 
circumstellar, or interstellar component, but it is sensitive to bulk gas 
motions. Excitation of the line is also sensitive to the ultraviolet photon 
environment. However, models made by Pasquini et al. (2011) and Dupree \& 
Avrett (2013) demonstrate utility as a He abundance indicator for globular 
cluster giants. 

Some empirical results suggest that the $\lambda$10830 \ion{He}{1} line may be 
informative to the debate over whether He inhomogeneities exist within globular
clusters. In the case of $\omega$ Centauri, which exhibits a large 
inhomogeneity in [Fe/H], Dupree et al. (2011) used the Phoenix spectrometer on 
the GEMINI-S telescope to discover a variation in $\lambda$10830 \ion{He}{1} 
line strength among 12 red giants chosen to occupy a limited region of the 
color-magnitude diagram. Although the helium transition was not detected in the
most metal-poor population (${\rm [Fe/H]} < -1.8$), it was identified in the
majority of stars with ${\rm [Fe/H]} \geq -1.8$. The appearance of helium 
correlated more closely with increased [Al/Fe] and [Na/Fe] abundances than with
[Fe/H]. This is analogous to what would be evinced as a CN-He correlation in a 
more usual bimodal-CN globular cluster. 

Dupree \& Avrett (2013) constructed semi-empirical chromosphere models of a 
pair of giants in $\omega$ Cen with strong and weak He lines and derived a 
difference of $\Delta Y \geq 0.17$ in the He mass fraction. They concluded that
the 1.083 $\mu$m \ion{He}{1} line can be used to determine a helium abundance 
for a cool star through the use of semi-empirical model chromospheres that are 
adequately constrained by other spectroscopic features such as the Balmer and 
\ion{Ca}{2} K lines. Furthermore, Pasquini et al. (2011) had earlier found that
the \ion{He}{1} line provided evidence of a He abundance difference between a 
pair of Na-rich and Na-poor giants in the cluster NGC 2808.

An example of a globular cluster with pronounced carbon and nitrogen abundance
inhomogeneities is the system of Messier 13. Extending to extremely blue 
colors, the horizontal branch of M13 suggests it as a particularly useful test 
case for He self-enrichment (e.g., Caloi \& D'Antona 2005). Central to the 
present paper is the question of whether CN-strong giants in a particular 
region of the RGB exhibit systematically stronger \ion{He}{1} lines than 
CN-weak giants. This paper reports upon the \ion{He}{1} transition of a sample 
of CN-strong and CN-weak red giants in M13 that has been observed with the 
NIRSPEC spectrometer on the Keck 2 telescope. 

In our previous NIRSPEC studies of the \ion{He}{1} line in metal-poor giants
(Smith et al. 2004; Dupree et al. 2009) the objective has mainly been to search
for asymmetries in the line profile indicative of mass motions and outflows. By
contrast, the emphasis in this paper is on determining whether CN-strong giants
in M13 have greater helium line strengths, and so it is the equivalent width 
that is of primary interest. Previous work on metal-poor field giants of 
metallicities similar to M13 has returned equivalent widths in the range 
0-100 m\AA\ (Dupree et al. 2009). 

\section{Observations and Spectra}

Spectra were acquired with the NIRSPEC spectrometer (McLean et al. 1998) on 
the Keck 2 telescope of seven red giants in Messier 13 along with two
giants in the cluster Messier 22. The general nature of the observing run was 
analogous to the NIRSPEC \ion{He}{1} programs reported by Smith et al. (2004) 
and Dupree et al. (2009).

Smith \& Briley (2006) published a compendium of CN-strong and CN-weak 
giants in Messier 13 based on data from the literature. Program stars  
were selected from their compilation in such a way as to include both CN-strong
and CN-weak giants of comparable luminosities. They are listed in Table 1
together with the visual magnitudes and $(B-V)$ colors tabulated by
Smith \& Briley (2006). The original star identifications are given in
Arp (1955) and Sandage (1970). Following Pilachowski et al. (1996) we adopt an
apparent distance modulus of $(m-M)_V = 14.33$ for M13, with the corresponding 
absolute magnitudes $M_V$ for the observed stars being listed in Table 1. 
Also tabulated are the values of a CN index denoted $m_{CN}$ which quantifies
the strength of the $\lambda$3883 CN band. The values are as given by Smith \& 
Briley (2006) while the index itself is defined by Suntzeff (1981).

Ancillary abundance information for the M13 stars observed is listed in 
Table 2. Column (2) gives the value of a CN-excess index $\delta m_{CN}$ that
is defined by Smith \& Briley (2006) and is based on making an empirical
correction to the $m_{CN}$ index to take account of the sensitivity of the
$\lambda$3883 CN band to stellar temperature and gravity. It is normalized so
that giants with the strongest CN bands have $m_{CN} > 1.0$. On the basis of 
$\delta m_{CN}$ the seven stars are classified as CN-strong, CN-intermediate, 
or CN-weak (denoted by an ``s'', ``i'', and ``w'' respectively in Tables 1 
and 2). Columns 4 and 5 of Table 2 give the values of [Na/Fe] and [O/Fe] that
were compiled by Smith \& Briley (2006; their Table 1) based in turn on
high resolution spectroscopic analyses by Sneden et al. (2004), 
Pilachowski et al. (1996) and Cohen \& Mel\'{e}ndez (2005). In the case
of the [O/Fe] abundances all of the measurements in column 5 come from
Sneden et al. (2004). Smith \& Briley (2006), building on previous work 
by Sneden et al. (2004) and Kraft et al. (1992, 1993, 1997), showed that 
the $\lambda$3883 CN band strength correlates well with the [Na/Fe] abundance 
in M13, while also anticorrelating with the [O/Fe] abundance, albeit with 
some scatter. Thus, use of the CN band strength should be a valid way of 
also picking out stars with different O and Na variations within the cluster.

A more recent study of sodium and oxygen abundances in M13 is that of
Johnson \& Pilachowski (2012), which includes 5 of the stars in our
NIRSPEC program. The [Na/Fe] and [O/Fe] abundances from Johnson \& 
Pilachowski (2012) are listed in columns 6 and 7 of Table 2. On the
basis of location within their plot of [Na/Fe] versus [O/Fe], Johnson
\& Pilachowski (2012) divided the M13 giants into three groups:
primordial (P), intermediate (I) and extreme (E), adopting the O/Na
classification scheme of Carretta et al. (2009). Despite the large
range in CN strengths all of the stars in our program fall into their
intermediate I category, which denotes ${\rm [Na/Fe]} \geq 0.0$ and
${\rm [O/Fe]} \geq -0.2$. In Appendix A the sodium and oxygen data
of Johnson \& Pilachowski (2012) are used to further study the
relationship between the abundance inhomogeneities of these elements 
relative to the $\lambda$3883 CN band strengths of red giant stars 
in the M13 cluster.

The two stars in Messier 22 for which spectra were obtained 
(when M13 was out of range of the Keck 2 telescope) are
numbered 69 and 87 in ring 1 of the color-magnitude study of Alcaino 
(1977), who measured their apparent magnitudes to be $V = 12.64$ and 12.48 
respectively. With an apparent distance modulus of $(m-M)_V = 13.74$ for M22 
(Monaco et al. 2004), the absolute magnitudes are $M_V = -1.10$ and $-1.26$ 
respectively.

Spectra were obtained on the nights of 2012 June 12 and 13 UT. The NIRSPEC 
spectrometer can record wavelengths in the range 0.95-5.4 $\mu$m via various
settings in either a high or a low resolution mode. Observations for our 
program were made with the high-resolution cross-dispersed echelle mode of 
NIRSPEC, using the NIRSPEC-1 blocking filter (bandpass 0.95-1.12 $\mu$m) 
to transmit those orders in the vicinity of the $\lambda$10830
\ion{He}{1} line. An echelle/cross-disperser combination of 
62.9/34.96 was used to place the order containing the \ion{He}{1} line in a 
suitable location on the detector. A slit of 0.43$\times$12 arc sec was
employed. NIRSPEC has an ALADDIN-2 1024$\times$1024 InSb array detector with 
27 $\mu$m pixels, that was typically used in the MCDS readout mode for the 
globular cluster stars. The standard pixel scale along the dispersion direction
in the high resolution mode is 0.144 arc sec pixel$^{-1}$, and the 0.43 arc sec
slit gives a resolving power of $\sim$ 24,000. Exposures were made as either 
AB nod pairs or in an ABBA nod pattern. Integration times at each position 
within a nod pattern typically ranged from 1200 s to 1800 s. Table 1 lists
the total integration time (in seconds) accumulated for each star.
The spectrograph slit was oriented along the parallactic angle
appropriate to each observation so that atmospheric dispersion of the starlight
was along the length of the slit. 

Spectra of the hot stars $\sigma$ Her and 13 Del were obtained to document 
telluric absorption over the range of airmass (1.0-1.6) at which the cluster
stars were observed. These spectra showed that the radial velocities of both
M13 ($-240$ km s$^{-1}$) and M22 ($-146$ km s$^{-1}$) place the \ion{He}{1}
$\lambda$10830 line far from any telluric absorption.
Flat field frames were acquired at the beginning and/or end of each night as 
well as on several occasions during the two nights when the NIRSPEC server and
consequently the NIRSPEC configuration had to be reset.
Arc spectra of a XeArNe comparison lamp were also obtained at the beginning 
and/or end of each night.

Reduction of the NIRSPEC images was accomplished by using the REDSPEC 
software package to produce stellar spectra of the order containing the 
$\lambda$10830 \ion{He}{1} line. Initial wavelength calibrations
were based on the XeArNe lamp spectra, with these solutions being shifted to 
match the rest frame of each cluster star. (Each spectrum was shifted
so as to place a strong absorption line of \ion{Si}{1} at the appropriate
rest wavelength of 10827 \AA). Spectra were normalized using
the IRAF\footnote{IRAF is written and supported by the National Optical
Astronomy Observatories (NOAO) in Tucson, Arizona. NOAO is operated by the
Association of Universities for Research in Astronomy (AURA), Inc. under
cooperative agreement with the National Science Foundation.} 
{\it continuum} task and fitting the stellar continuum with
a 5th order cubic spline.

Counts in the spectra in the vicinity of the $\lambda$10830 \ion{He}{1} line 
range from 2800 ADU per pixel for star IV-19 to 10000 ADU per pixel for star 
J3. The gain of the detector was 5.7 e$^{-}$ per ADU. In the region of the 
\ion{He}{1} line the pixel scale is 0.156 \AA\ per pixel. 

A montage of the M13 spectra is shown in Fig.~1 with the wavelength scale in 
each panel being that of the rest frame of the designated star. Due to the 
large negative radial velocity of M13 there are no telluric lines of any 
significance in the wavelength region covered by Fig.~1. Within the NIRSPEC 
spectra the \ion{He}{1} triplet line of the M13 giants typically has the form 
of a shallow and broad feature. Located about 3 \AA\ to the blue of the 
\ion{He}{1} feature is the strong \ion{Si}{1} transition. Six out of the 
seven M13 giants observed show a \ion{He}{1} line absorption profile. However, 
for star X24 there is emission at the rest wavelength of the \ion{He}{1} line 
which appears to show a P Cygni profile with absorption blueward of 
10,828.5 \AA.

The pair of giants observed in M22 show very shallow \ion{He}{1} absorption
and \ion{Si}{1} lines having comparable depths to those of the M13 giants.
Spectra of these two stars are presented in Fig.~2. The large negative radial 
velocity of M22 again has a consequence that there are no telluric lines of 
any significance in the wavelength region covered by Fig.~2.

Measurements of the equivalent width $EW$ of each \ion{He}{1} line were 
attempted in two ways from continuum-normalized spectra. In one approach the 
equivalent widths were measured with the IRAF task {\it splot} using an 
integration of the observed counts over the wavelength extent of the absorption
line as judged by eye. The helium line is shallow and broad, and upon doing a 
direct integration it is difficult to accurately account for the extent to 
which the blue wing of the \ion{He}{1} line may blend into the $\lambda$10827
\ion{Si}{1} line. The direct integration procedure involved marking the two 
limits of the \ion{He}{1} line where the profile reaches the continuum, and 
then integrating the data between the two points. In order not to include the
\ion{Si}{1} line in this approach, the limit on the short wavelength side of 
the \ion{He}{1} line was chosen so as not to intrude upon the red side of the 
\ion{Si}{1}. This approach potentially underestimates the equivalent width of 
the \ion{He}{1} line, which visually seems to be quite broad for many of the 
M13 giants. As a second approach multiple fits using the IRAF {\it splot} 
deblending tool were made to the \ion{Si}{1} and \ion{He}{1} lines assuming 
that both had Gaussian profiles. The deblending procedure fits two Gaussians 
simultaneously, one centered on the \ion{Si}{1} line, and the other on the 
\ion{He}{1} line. This second approach, which attempts to correct the He line 
$EW$ for blending with the Si line profile, gave greater \ion{He}{1} equivalent
widths than the first approach, but with the added uncertainty that the He line
may not be Gaussian in shape or symmetrical. As a consequence both sets of 
\ion{He}{1} equivalent widths, denoted $EW_1$ and $EW_2$ respectively according
to approaches 1 and 2, are listed in Table 1, along with the equivalent width 
of the \ion{Si}{1} line from the Gaussian fit.

A test was done to ascertain how an error in the continuum fitting can
transform into an error in the equivalent width. The value of $EW_1$ for the 
\ion{He}{1} line was remeasured for three stars after adjusting the continuum 
level slightly and re-integrating over the line. Changes in the measured 
equivalent widths spanned 7 to 12\% from the average equivalent width, and we 
adopt a value of 10\% as a reasonable assessment of the uncertainty in 
$EW_1$ due to continuum placement errors. We note that for some stars the 
difference between the values of $EW_1$ and $EW_2$ exceed this error estimate 
from the $EW_1$ measurements. However, as noted above, the short wavelength 
wing of the helium line may overlap the wing of the \ion{Si}{1} feature, so 
that the measured value of $EW_1$ is in some cases a lower 
limit to the actual equivalent width. 

The star X24 shows a P Cygni profile with extended emission and a weak 
absorption component. Consequently this CN-strong giant does not provide a 
suitable \ion{He}{1} comparison with the other six M13 giants observed, and 
values are not listed for either $EW_1$ or $EW_2$ in Table 1.
  
\section{Results}

\subsection{The $\lambda$10830 \ion{He}{1} Line Equivalent Width}

Equivalent widths $EW_1$ for the \ion{He}{1} line are plotted versus the
CN index $\delta m_{CN}$ in Fig.~3 for the Messier 13 giants, with filled
circles denoting CN-strong stars and open circles depicting either
CN-weak or CN-intermediate giants. Conspicuous by its absence is any 
indication that the CN-weak giants have weaker \ion{He}{1} lines than the 
CN-strong stars. Not represented in Fig.~3 due to the P Cygni nature of 
its \ion{He}{1} feature is the CN-strong giant X24.

The alternative set of equivalent widths $EW_2$ is shown versus $\delta m_{CN}$
in Fig.~4. As with the preceding figure there is no evidence of a correlation 
between CN band strength and He line equivalent width. Thus despite 
uncertainties over the appropriate technique for measuring $EW$(\ion{He}{1})
the conclusion of this paper is that of six M13 red giants there is no 
evidence of a He-CN correlation, i.e., no evidence that the CN-strong
giants have systematically stronger \ion{He}{1} lines.

Using sodium or oxygen abundances in place of $\delta m_{CN}$
does not produce any notable trends.\footnote{There may be a visual
impression of an anticorrelation between $EW_{1,2}$(He) and CN-band strength
in Figs.~3 and 4, or between $EW_1$(He) and [Na/Fe] in Fig.~5. We are
reluctant to attach any significance to these possibilities. In both
Figs.~5 and 7 if the one data point with the largest value of $EW_1$
was taken away, or the value of $EW_1$ reduced by $\sim$ 10\%, the
appearance of an anticorrelation would be greatly diminished. We prefer
instead to stress that the data yield no indication of correlations between
$EW$(He) and either CN band strength or Na abundance.} Plots of $EW_1$ 
versus [Na/Fe] and [O/Fe] from columns 4 and 5 of Table 2 are shown in 
Figures 5 and 6. There is no indication of either a correlation between 
He $EW_1$ and [Na/Fe], or an anticorrelation between He $EW_1$ and [O/Fe]. 
Using the [Na/Fe] and [O/Fe] abundances of Johnson \& Pilachowski (2012) 
would lead to fewer stars in each of Figs.~5 and 6 but no difference in 
the conclusions.

Consequently, among the stars in our NIRSPEC program there is no evidence 
that those with strong CN bands and enhanced Na, but comparatively low oxygen 
abundances, have stronger $\lambda$10830 \ion{He}{1} lines than stars of 
weak CN, low [Na/Fe], and high [O/Fe]. Whatever primordial enrichment process 
produced the high sodium but reduced oxygen abundances of the CN-strong 
stars in Table 1, there is no indication that this process was accompanied by 
enrichment in helium. Whether there may be some helium-enriched stars in 
Messier 13 that have eluded our modest sample is discussed in Sec 3.4.

\subsection{The $\lambda$10830 \ion{He}{1} Line Profile}

Visual comparison of the spectra themselves reveal no consistent pattern
in \ion{He}{1} lines. Six of the stars in Table 1 can be conveniently divided
into three CN-strong/CN-weak (or CN-intermediate) pairs (J3/A1, I-12/IV-19, 
I-18/II-41) on the basis of their position in the color-magnitude diagram 
(CMD) as shown in Fig.~7. The giants within each of these three pairs have 
very similar $V$ magnitudes and $B-V$ colors, with each pair being somewhat 
separated in the CMD. The colors are corrected for a reddening of $E(B-V)=0.02$
(Harris 1996). The dashed line represents a fiducial locus for the M13 red 
giant branch taken from Sandage (1970).

Spectra for the stars in each CN-pair are shown superimposed in Fig.~8.
In the J3/A1 pair there is little difference in the $\lambda$10830 feature
between the two stars, while in the case of the I-12/IV-19 pair the CN-strong
star I-12 exhibits a weaker He feature. Only among the I-18/II-41 pair does
the CN-strong star possibly show slightly more absorption at the rest frame
wavelength of the He feature. Figure 8 displays in a visual way, independent of
equivalent width measurements, our main finding that the spectra provide no 
indication that the CN-strong stars in our M13 sample have systematically 
stronger $\lambda$10830 \ion{He}{1} lines than the CN-weak giants.

The $\lambda$10830 \ion{He}{1} line of the seven M13 giants observed shows a 
range of profiles. It exhibits for some stars an asymmetric profile that 
extends blueward towards the Si line. As with the previous studies of the 
\ion{He}{1} line among metal-poor red giants by Smith et al. (2004) and 
Dupree et al. (2009), there is evidence for some stars in the M13 sample of 
chromospheric outflows that have imposed extended absorption upon the blue 
wing of the line. 

At the wavelength of the $\lambda$10830 \ion{He}{1} line a radial velocity of
$-50$ km s$^{-1}$ corresponds to a blue shift in the spectrum of 1.8 \AA. In 
order for the \ion{He}{1} line to exhibit blueward absorption extending to the 
vicinity of the $\lambda$10827 \ion{Si}{1} line it would be necessary for the
chromosphere to contain an outflow whose velocity field reaches 
$\sim$ 80 km s$^{-1}$. Short-wavelength absorption in the \ion{He}{1} profile
that extends to velocities of this order seems to be present for the stars 
II-41 and IV-19, the P Cygni star X24, and possibly A1 and I-18. These
broadened profiles indicate that the He line is being substantially
influenced by the velocity fields of the outer chromospheres of M13
red giants, at least those with absolute magnitude brighter than $M_V = -0.3$.
It is worth noting that Suntzeff (1981) considers X24 to be a first-ascent red
giant branch star; the P Cygni nature of the \ion{He}{1} line is not 
attributable to an AGB star. 
 
If the \ion{He}{1} line is to become a potential tool for He abundance 
determinations in metal-poor red giants of the Galactic halo, it would seem
that mass motions in the chromosphere must be incorporated into the models
to derive abundance information.

M\'{e}sz\'{a}ros et al. (2009) studied the asymmetries of both H$\alpha$
and \ion{Ca}{2} K emission line profiles among red giants in M13. In their
sample those red giants with evidence of outflows in either 
emission profile have absolute magnitudes $M_V$ brighter than $-1.4$ mag.
Another sign of outflow is the velocity shifted core of the H$\alpha$ line
which is formed higher in the chromosphere than wing emission. In the
M\'{e}sz\'{a}ros et al. (2009) sample H$\alpha$ core outflow occurs for
stars with $\log (L/L_{\odot}) \geq 2.5$, which for M13 corresponds to
stars brighter than $M_V \sim -0.9$. As such the evidence for outflows from 
the \ion{He}{1} line extends to less luminous giants than those traced by 
the H$\alpha$ and \ion{Ca}{2} K line observations.

\subsection{Contrasting M13 and $\omega$ Centauri}

The \ion{He}{1} equivalent widths presented here for the M13 giants can be
compared with those found by Dupree et al. (2011) for $\omega$ Centauri.
Of the M13 stars in Table 1 there are several with values of $EW$(He) 
less than 40 m\AA, whereas for the detections in $\omega$ Cen the typical 
values are greater than 50 m\AA\ and extend to $\sim$ 200 m\AA. Thus Dupree 
et al. (2011) find some giants in $\omega$ Cen with much stronger
$\lambda$10830 \ion{He}{1} lines than are present among either the CN-weak or
CN-strong giants of M13 listed in Table 1. So not only is there a lack
of a CN-He correlation in M13, but the distribution of $\lambda$10830
$EW$s is different from that in $\omega$ Cen, and more like that of the
Population II field stars in Dupree et al. (2009).

Studies of the horizontal branch morphology of M13 suggest that the helium
mass fraction may be as high as $Y \sim 0.30$ among some stars (Dalessandro
et al. 2013), and 0.02-0.04 higher than for the comparable metallicity cluster
M3. Sandquist et al. (2010), on the other hand, find that the $R$ ratio of
M13, a measure of the relative numbers of red giants and horizontal branch 
stars (Cassisi et al. 2003), is consistent with a helium mass fraction of 
$Y=0.25$, with no evidence for a $\Delta Y \sim 0.04$ enhancement above 
a standard Big Bang level. By contrast, much larger helium contents of
$Y \sim 0.4$ (Norris 2004; Piotto et al. 2005; King et al. 2012;
Dupree \& Avrett 2013) have been invoked or derived for some stars in 
$\omega$ Centauri. Consequently, the current finding that the $\lambda$10830
\ion{He}{1} feature equivalent widths among red giants in M13 do not extend
to the larger values found by Dupree et al. (2009) for $\omega$ Cen would 
seem at least qualitatively consistent with expectations based on the 
morphologies of the color-magnitude diagrams of these two clusters.

\subsection{A Direction for Further Work: Are Helium Enhancements Hiding Among 
the Most Oxygen-Depleted Giants in M13?}

The lowest oxygen abundance among the stars in our NIRSPEC sample is somewhat 
uncertain. On the basis of the O abundances tabulated in column 4 of Table 2
the most oyxgen-poor star for which we have a \ion{He}{1} spectrum is I-12, 
the original [O/Fe] measurement of Sneden et al. (2004) being $-0.47$ dex.
If converted to the oxygen abundance scale of Johnson \& Pilachowski (2012), 
the Sneden et al. abundance for star I-12 would be ${\rm [O/Fe]} \sim -0.38$
(see Appendix 1). However, Johnson \& Pilachowski (2012) derive a much 
higher oxygen abundance for this star, finding it to be similar in [O/Fe] 
to J3. There is no [O/Fe] measurement in the literature for I-18, the third 
CN-strong stars in our NIRSPEC sample. 

There are red giants in M13 that are more oxygen-poor than star I-12. Following
Johson \& Pilachowski (2012), it remains a possibility that He enhancements in 
M13 might be confined to those red giants of the lowest oxygen abundance, 
namely stars in the E category of Johnson \& Pilachowski (2012) for which 
${\rm [O/Fe]} < -0.2$. The oxygen abundance of star I-12 found by Sneden et al.
(2004) does place it in the E category, but not the result of Johnson \& 
Pilachowski (2012). Regardless of this uncertainty as regards I-12, there are 
nonetheless red giants of M13 in both the studies of Johnson \& Pilachowski 
(2012) and Sneden et al. (2004) that have oxygen abundances of 
${\rm [O/Fe]} < -0.4$. It would seem worthwhile extending \ion{He}{1} 
observations to the most oxygen-poor giants known in M13. 

Interpreting the behavior of the $\lambda$10830 \ion{He}{1} line among the most
oxygen-deficient giants in M13 may hold challenges. The work of Kraft et al.
(1993, 1997) indicated that the very lowest [O/Fe] values within M13 tended to
be concentrated among the most luminous giants in the cluster, a result which
suggested to them that stellar interior mixing on the upper red giant branch
has contributed to the great oxygen depletions. The results of Johnson \& 
Pilachowski (2012) seem to support this, as they themselves stress for
several reasons: (i) all known stars with ${\rm [O/Fe]} < -0.4$ have 
luminosities of $\log (L/L_{\odot}) > 2.0$, (ii) among the E population of
giants in M13 the average [O/Fe] abundance decreases with increasing
luminosity, and (iii) among the highest luminosity giants in the E and I 
categories there are large differences in [O/Fe] but only small differences
in [Na/Fe]. Smith et al. (2004) found that the most luminous metal-poor red 
giants in their field and M13 sample often did not exhibit $\lambda$10830 
\ion{He}{1} absorption features. It was this finding that lead us to 
concentrate our newer NIRSPEC observations on stars in M13 that are well 
removed in absolute magnitude from the tip of the red giant branch, in an 
effort to enhance the likelihood of detecting the \ion{He}{1} line. This 
selection, however, drove our program away from the upper reaches of the 
red giant branch in M13 where the most oxygen-deficient giants seemingly 
prefer to reside. 

Thus, one avenue for further $\lambda$10830 \ion{He}{1} observations would be 
to focus on the faintest very-oxygen-poor giants in M13, in an effort to 
determine whether He inhomogeneities occur among those cluster stars with the 
most extreme O-Na-CN combinations. Such red giants, however, with 
$\log (L/L_{\odot}) > 2.0$, may have had their oxygen abundances diminished 
below any primordial (pre-red-giant) 
level by interior deep mixing, i.e., any He enhancements that are found might 
not be unambiguously attributable to enrichment of M13 by early generations of 
intermediate-mass or high-mass stars. The current NIRSPEC sample of Table 1 
does encompass a substantial range in CN band strengths, O and Na abundances, 
while avoiding the most oxygen-depleted giants that might be complicated by 
(non-convective) deep mixing. As such they depict a reasonable subset of the 
abundance inhomogeneities in M13, and as such they reflect ``typical'' levels 
of primordial enrichment in that cluster. For example, Johnson \& Pilachowski 
(2012) found that giants in their I category constitute almost 65\% of the 
stars in M13, while very-oxygen-poor giants with ${\rm [O/Fe]} < -0.2$ comprise
$\sim$ 22\% of the cluster. Thus the indication of our present NIRSPEC 
spectroscopy is that He enhancements do not appear to be ``typical'' products 
of the primordial enrichment of M13. Observations of the most O-poor giants in
M13 are needed to determine whether He enhancements might be present among
the most extreme stars in that cluster. 

\subsection{M13 IV-15}

Several red giants in M13 were observed in the \ion{He}{1} NIRSPEC program of 
Smith et al. (2004). The stars observed by them were reasonably luminous, and 
high up on the red giant branch, brighter than those reported upon in this 
paper. Some of their M13 stars did not show a \ion{He}{1} absorption line. One 
star that did is number IV-15 in the photographic study of Arp (1955). It is
is a CN-weak star (Smith \& Briley 2006) with an apparent magnitude of
$V = 12.96$, making it more than 0.4-1.0 mag brighter than the stars in 
Table 1. The equivalent widths of the \ion{He}{1} line were measured from this 
earlier spectrum and found to be $EW_1 = 101$ m\AA\ and $EW_2 = 123$ m\AA, 
which place it at the upper bound of the line strengths of the stars in 
Table 1. As such, IV-15 seemingly provides no evidence for a positive 
correlation between CN band strength and He line strength. However, Smith \& 
Briley (2006) classify IV-15 as an asymptotic giant branch star with 
$(B-V) = 1.04$, and as such it does not make for a straightforward comparison 
with the M13 red giant branch stars in Table 1 due to the different 
temperature-gravity relations between the red giant and asymptotic giant 
branches. Johnson \& Pilachowski (2012) measured an oxygen abundance
for IV-15 of ${\rm [O/Fe]} = +0.10$, so it does not appear to be an 
example of an oxygen-poor star. Both Johnson \& Pilachowski (2012) and 
Sneden et al. (2004) find a sodium abundance of ${\rm [Na/Fe]} \sim +0.1$ 
for this star, which is mid-range for the spread observed in M13. Consequently,
IV-15 does not appear to be a candidate for an oxygen-poor but CN and He 
enhanced star of the type that is missing from the sample in Table 1.

\subsection{The M22 Pair}

Designated as stars I-53 and I-85 in the color-magnitude study of Arp \&
Melbourne (1959) the giants referred to in Fig.~2 as Alcaino 1069 and 1087
respectively were included in the abundance study of M22 made by Marino
et al. (2011). There is a considerable difference in both oxygen and
sodium abundances between the stars in this pairing. The more sodium-rich
star is Alc 1087 with ${\rm [Na/Fe]} = 0.59$ compared to 0.29 dex for
Alc 1069. There is a sodium-oxygen anticorrelation between these giants, with
a difference in [O/Fe] of 0.38 dex. Alc 1087 with an oxygen abundance of
${\rm [O/Fe]} = -0.03$ is one of the most oxygen-poor stars in the Marino et
al. (2011) study, within which the lowest oxygen abundance encountered is
${\rm [O/Fe]} = -0.10$. Marino et al. (2011) and Roederer et al. (2011) found 
inhomogeneities in {\it s}-process elements in M22, with Alc 1069 being a 
member of the {\it s}-rich subpopulation and Alc 1087 being {\it s}-poor. There
is also an inhomogeneity in iron and calcium within M22 (Norris \& Freeman 
1983; Lehnert et al  1991; Anthony-Twarog et al. 1995; Da Costa et al. 2009;
Marino et al. 2011; Da Costa \& Marino 2011),
however the difference in [Fe/H] between Alc 1069 ($-1.74$) and Alc 1087
($-1.81$) is modest. Comparing the spectra of these two stars in Fig.~2
reveals a slightly stronger \ion{He}{1} feature for Alc 1087, the more
oxygen deficient of the pairing. There may be a suggestion here of 
a modest He-O anticorrelation and He-Na correlation. As such, M22 would be
worth further investigation as a possible candidate for He inhomogeneities
linked to those of the elements oxygen and sodium.  
 
\section{Summary}

Whereas inhomogeneities in the abundances of certain elements such as nitrogen
are commonplace among the red giants of M13, the $\lambda$10830 \ion{He}{1} 
spectra presented in this paper do not provide compelling evidence for
correlated He enhancements in stars with a strong $\lambda$3883 CN band, which
is a surrogate for the nitrogen abundance. Whereas there is considerable 
evidence in the literature that heterogeneous primordial enrichment of 
globular clusters contributed to the inhomogeneities in CN, O, and Na in 
clusters such as M13, there is no evidence from our NIRSPEC observations 
that these inhomogeneities extend to the element helium, with a caution
noted in the following paragraph.

Circumstantial evidence based on the morphology of the horizontal branch 
in the color-magnitude diagram suggests that any He inhomogeneities in 
M13 may be more modest than those derived for $\omega$ Cen from the 
\ion{He}{1} line by Dupree \& Avrett (2013). A caveat to our results is 
that $\lambda$10830 spectra have yet to be obtained for the most
oxygen-deficient giants in M13. Oxygen abundances of ${\rm [O/Fe]} < -0.2$ 
were found by Johnson \& Pilachowski (2012) among about 22\% of the red giants
in M13, and it is upon such stars that further spectroscopy of the
\ion{He}{1} line would be worth concentrating. Perhaps it is among such stars
that He abundance enhancements might be encountered, although the potential
occurrence of deep interior mixing within such stars could complicate the
interpretation of their origin.

Given that a number of the $\lambda$10830 \ion{He}{1}
absorption profiles among the M13 red giants show evidence of 
mass motions in the chromosphere, it could be challenging to detect modest He 
variations in the cluster via this technique. A detailed matching 
of the line profile computed for chromospheric models that incorporate mass
outflows would seem to be a necessary prerequisite for a measurement of the
He abundance via the $\lambda$10830 line, unless a large enough number of M13 
stars can be surveyed so as to obtain a sample in which the line shows
no evidence of asymmetric velocity broadening.

\acknowledgments

Thanks are extended to Dr.~Greg Wirth for serving as the support astronomer on
the Keck observing run during which the NIRSPEC observations for this paper
were acquired. The authors wish to recognize and acknowledge the very 
significant cultural role and reverence that the summit of Mauna Kea has always
had within the indigenous Hawaiian community. We are most fortunate to have the
opportunity to conduct observations from this mountain. Thank you to the
referee for a useful review of the manuscript. GHS gratefully
acknowledges the support of NSF award AST-0908757.

\appendix

\section{The CN, O, Na Inhomogeneities in Messier 13}

The abundance data from Johnson \& Pilachowski (2012; JP) allow us to revisit 
the relationships between CN band strength and [O/Fe] and [Na/Fe] studied 
earlier for M13 by Sneden et al. (2004; S04) and Smith \& Briley (2006; SB). 
In Fig.~9 the homogenized [Na/Fe] abundances tabulated by Smith \& Briley 
(2006) are plotted versus the Johnson \& Pilachowski (2012) determinations, 
while in Fig.~10 the [O/Fe] abundance data from Sneden et al. (2004) and 
Johnson \& Pilachowski (2012) are compared. In both figures a straight line
depicts equality between the two abundance sets.

There are 29 stars in common to the [Na/Fe] measurements of Johnson \& 
Pilachowski (2012) and the tabulation of Smith \& Briley (2006). The mean
difference is [Na/Fe](JP) $-$ [Na/Fe](SB) = $-0.04$ with a standard deviation
of 0.20 dex. Thus it seems as if the two [Na/Fe] abundance sets are on the
same system. By contrast, as regards the [O/Fe] data there are 22 stars in 
common between JP and S04, with a mean difference of [O/Fe](JP) $-$ [O/Fe](S04)
= 0.085 and a standard deviation of 0.21 dex. A dashed line has been added to
Fig.~10 to illustrate this mean difference.

Just how well the CN band strength of M13 giants traces the Na and O abundance
variations can be judged from Figures 11 and 12 in which the 
$\delta m_{CN}$ index is plotted versus [Na/Fe] and [O/Fe] respectively
from Johnson \& Pilachowski (2012). In accord with previous findings of
Kraft et al. (1993, 1997), Sneden et al. (2004) and Smith \& Briley (2006)
this pair of figures shows that CN band strength correlates with [Na/Fe]
abundance while anticorrelating with [O/Fe]. There is scatter in the 
relationships, for example, at a given CN band strength there may be a
range of $\sim \pm 0.2$ dex in [Na/Fe] among the CN-weak giants with 
$\delta m_{CN} < 0.5$, and a spread of $\sim \pm 0.1$ dex in [Na/Fe] among 
the CN-strongest giants with $\delta m_{CN} \geq 1.0$. Based on the scatter 
seen between the independent [Na/Fe] datasets in Fig.~11 it is possible that a 
considerable component of the scatter in [Na/Fe] among the CN-strongest giants
stems from measurement uncertainties. Among the CN-weakest giants in Fig.~11 
there may be intrinsic scatter in [Na/Fe] since Johnson \& Pilachowski (2012) 
assess the average measurement error in their [Na/Fe] derivations at 0.08 dex. 

There are seven stars in Fig.~12 that fall into the Extreme category having 
the lowest oxygen abundances (${\rm [O/Fe]} \leq -0.2$). All seven of these
stars are CN-strong with $\delta m_{CN} \geq 0.9$, and six out of seven 
have $\delta m_{CN} \geq 1.0$. Nonetheless there is one giant in Fig.~12 that
has both strong CN bands and a very high oxygen abundance of 
[O/Fe] $\sim$ +0.4, and several CN-strong giants have [O/Fe] close to the
solar ratio. Thus although there is on average an anticorrelation between
CN band strength and oxygen abundance there may be some intrinsic scatter
in the cluster of M13. The usual explanation of the trend seen in Fig.~12 
is that the material in the CN-strong giants of M13 has been subjected to 
the O$\rightarrow$N cycle of hydrogen burning (e.g., Wallerstein et al. 1987;
Brown et al. 1991; Kraft et al. 1992) 
relative to the material in the CN-weak giants. Johnson \& Pilachowski (2012)
discussed the alternative scenarios for how this may have come about through 
a combination of primordial enrichment of the young M13 protocluster plus 
interior mixing within current-day red giants during the most advanced stages 
of RGB evolution. Briley et al. (2002, 2004) and Smith \& Briley (2006) 
had previously discussed the evidence for such a dual-origin to
the CNO inhomogeneities in Messier 13. 

The relationships between CN, [Na/Fe], and [O/Fe] map into an anticorrelation
between [Na/Fe] and [O/Fe] within M13 that has been studied and commented upon
by Kraft et al. (1992, 1993, 1997), Sneden et al. (2004), Cohen \& 
Mel\'{e}ndez (2005), and Johnson \& Pilachowski (2012). A O-Na anticorrelation
has been extensively documented among both red giant branch and horizontal
branch stars in many other globular clusters of the Milky Way (e.g., 
Kraft et al. 1995, 1998; Sneden et al. 1992, 1997; Shetrone \& Keane 2000;
Ivans et al. 1999, 2001; Ram\'{i}rez \& Cohen 2002;
Carretta et al. 2006, 2007a,b,c, 2009, 2014; Gratton et al. 2011, 2012, 2013; 
Yong \& Grundahl 2008; Yong et al. 2005, 2014; and references therein).

\clearpage

\begin{figure}
\figurenum{1}
\begin{center}
\includegraphics[scale=0.7]{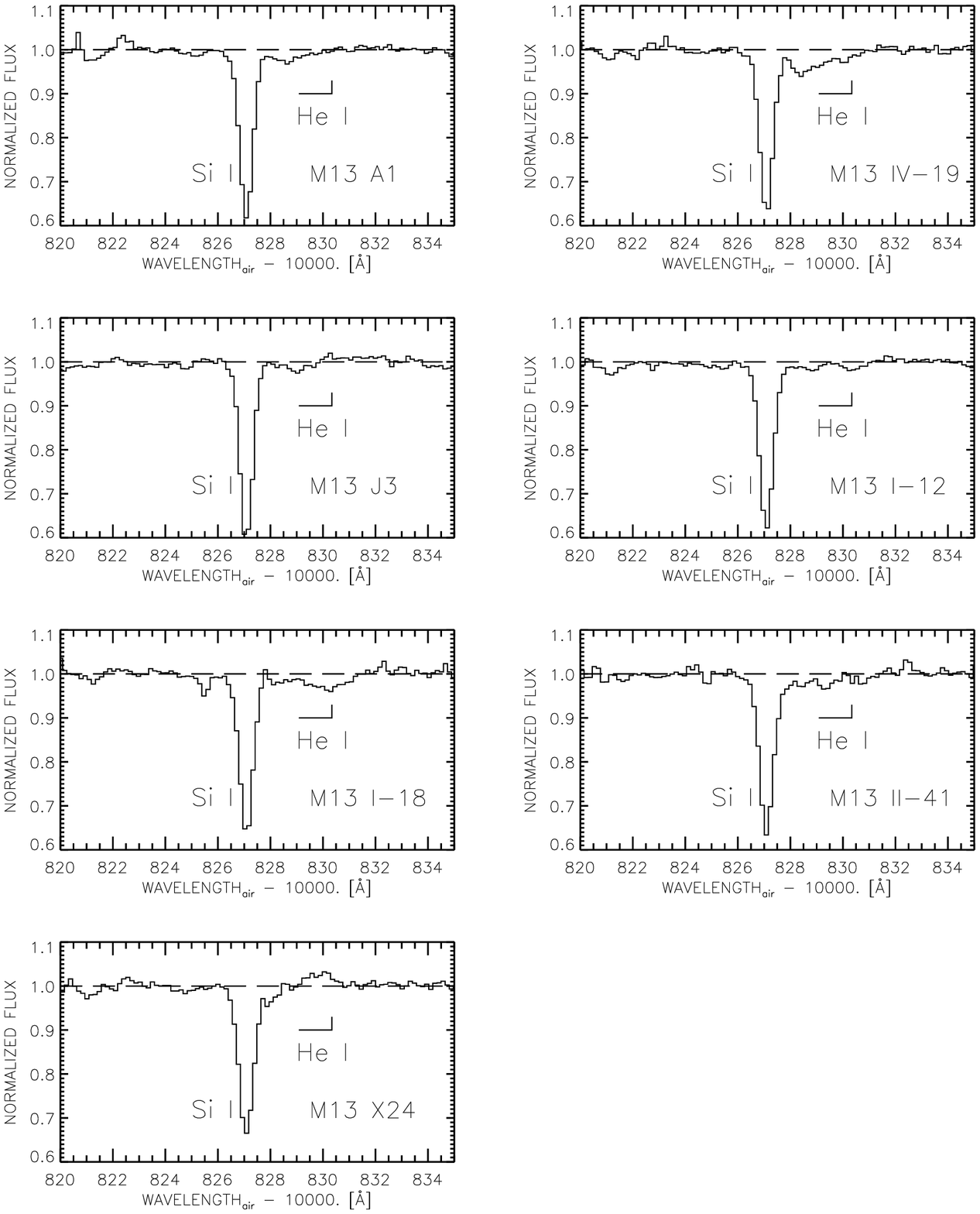}
\caption{Spectra in the vicinity of the $\lambda$10830 \ion{He}{1}
feature for seven red giants in the globular cluster Messier 13.}
\end{center}
\end{figure}
\clearpage

\begin{figure}
\figurenum{2}
\begin{center}
\includegraphics[scale=0.7]{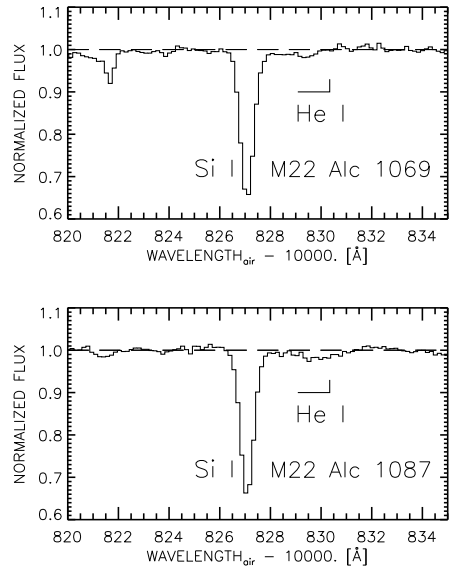}
\caption{Spectra in the vicinity of the $\lambda$10830 \ion{He}{1}
feature for two red giants in the globular cluster Messier 22.}
\end{center}
\end{figure}
\clearpage

\begin{figure}
\figurenum{3}
\begin{center}
\includegraphics[scale=0.7]{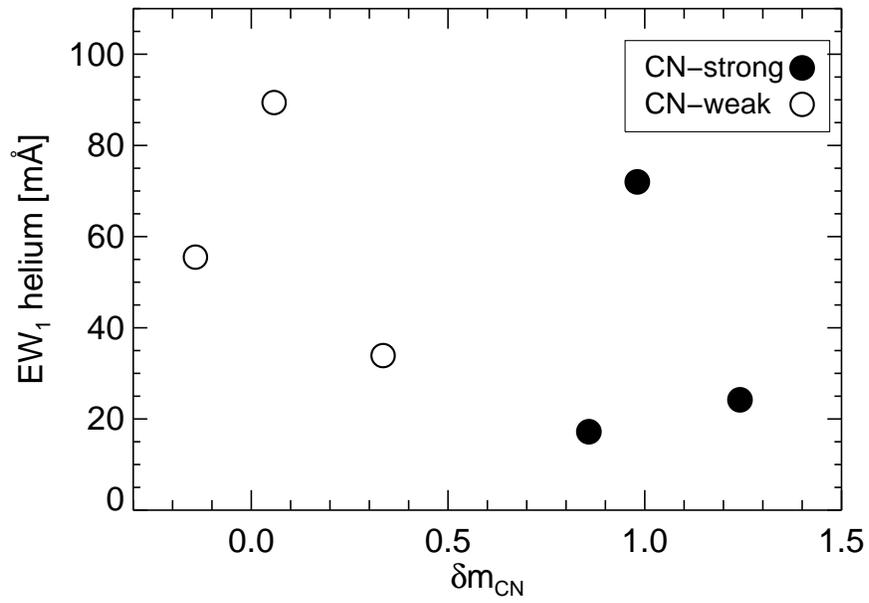}
\caption{The $EW_1$ measurement of the equivalent width of the \ion{He}{1} 
10830 \AA\ line versus the CN index $\delta m_{CN}$ for Messier 13 giants. 
Filled and open circles denote CN-strong and CN-weak/CN-intermediate giants 
respectively. Due to its \ion{He}{1} line having
a P Cygni profile the star X24 is not shown in the figure.}
\end{center}
\end{figure}
\clearpage

\begin{figure}
\figurenum{4}
\begin{center}
\includegraphics[scale=0.7]{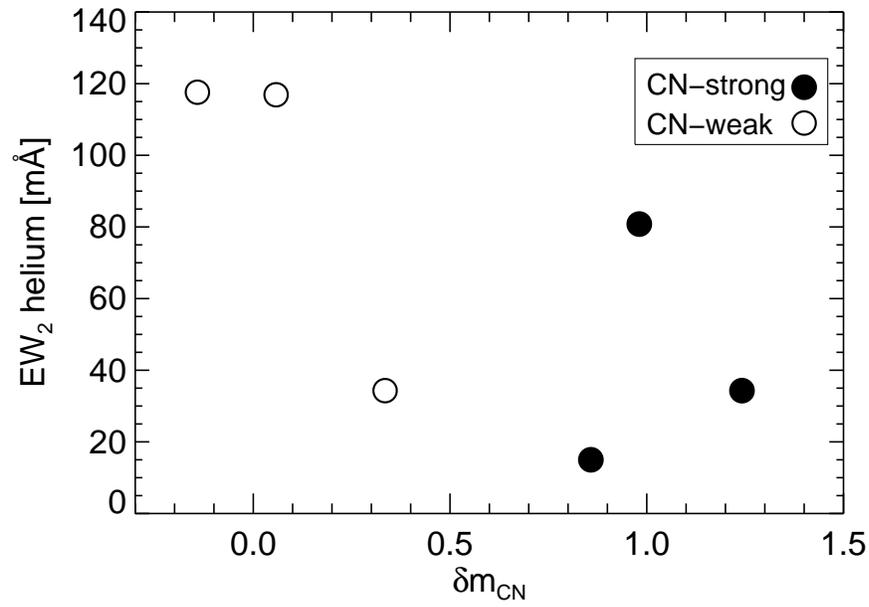}
\caption{The $EW_2$ measurement (deblended profile) of the \ion{He}{1} line 
equivalent width versus the $\delta m_{CN}$ CN index for Messier 13 giants. 
Filled and open circles denote CN-strong and CN-weak/CN-intermediate giants 
respectively. Star X24 is not shown.}
\end{center}
\end{figure}
\clearpage

\begin{figure}
\figurenum{5}
\begin{center}
\includegraphics[scale=0.7]{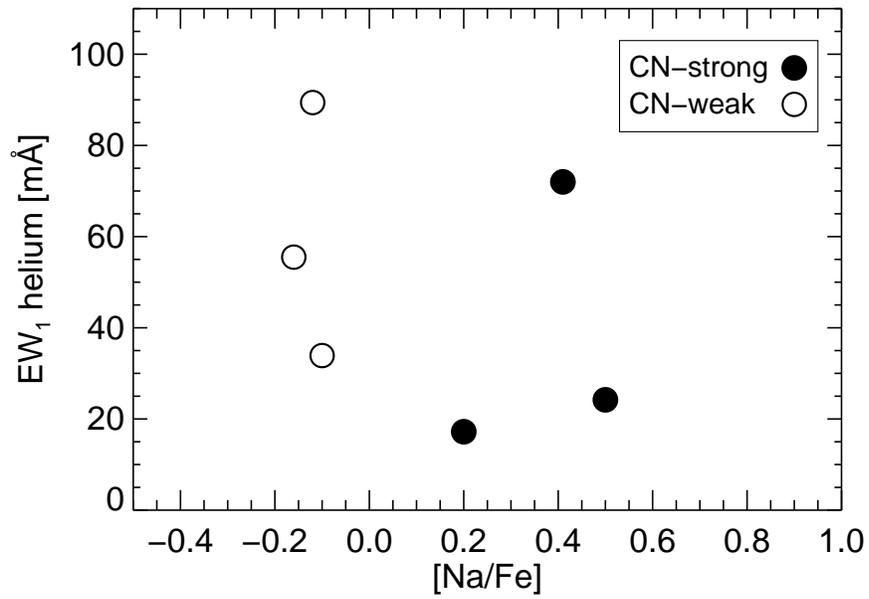}
\caption{The $EW_1$ measurement of the equivalent width of the \ion{He}{1} 
10830 \AA\ line versus [Na/Fe] abundance from column 5 of Table 2 for 
Messier 13 giants. Filled and open circles denote CN-strong and 
CN-weak/CN-intermediate giants respectively. Star X24 is excluded from the 
figure on account of exhibiting a P Cygni profile in the \ion{He}{1} line.}
\end{center}
\end{figure}
\clearpage

\begin{figure}
\figurenum{6}
\begin{center}
\includegraphics[scale=0.7]{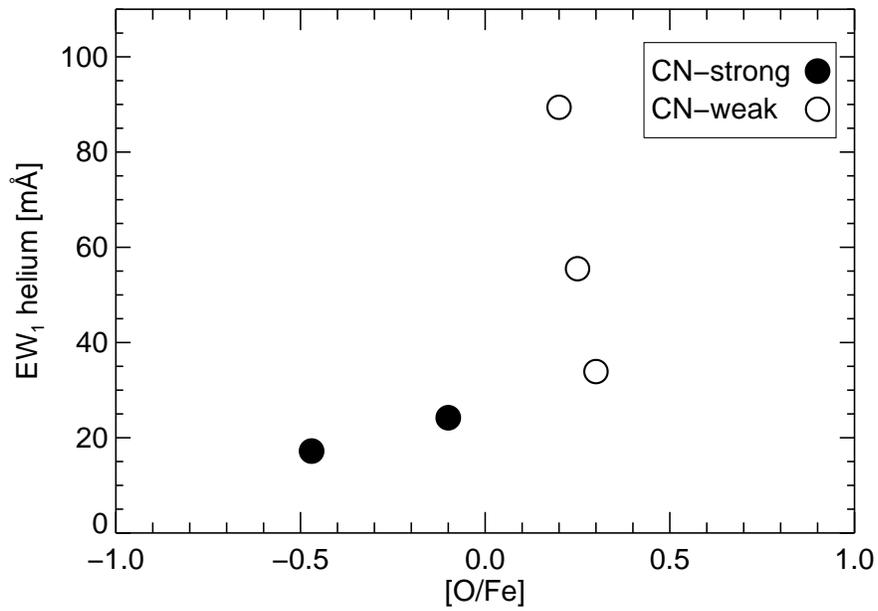}
\caption{The equivalent width $EW_1$ of the \ion{He}{1} 10830 \AA\ line versus 
[O/Fe] abundance from column 4 of Table 2 for M13 giants. Filled and open 
circles denote CN-strong and CN-weak/CN-intermediate giants respectively. 
Star X24 is again omitted from the plot due to a P Cygni profile
in the \ion{He}{1} line.}
\end{center}
\end{figure}
\clearpage

\begin{figure}
\figurenum{7}
\begin{center}
\includegraphics[scale=0.7]{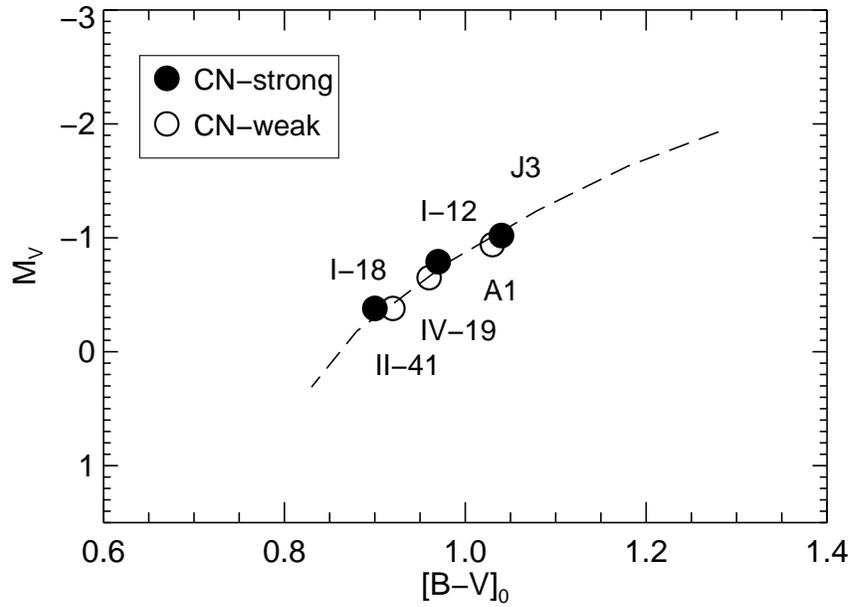}
\caption{The color-magnitude diagram of three pairs of Messier 13 giants
with NIRSPEC spectroscopy. Filled and open circles denote CN-strong and 
CN-weak/CN-intermediate giants respectively. Star X24 is not shown. Each point 
is labelled with the designation of each star, while the dashed line is based
on the fiducial red giant branch of Sandage (1970).}
\end{center}
\end{figure}
\clearpage

\begin{figure}
\figurenum{8}
\begin{center}
\includegraphics[scale=0.7]{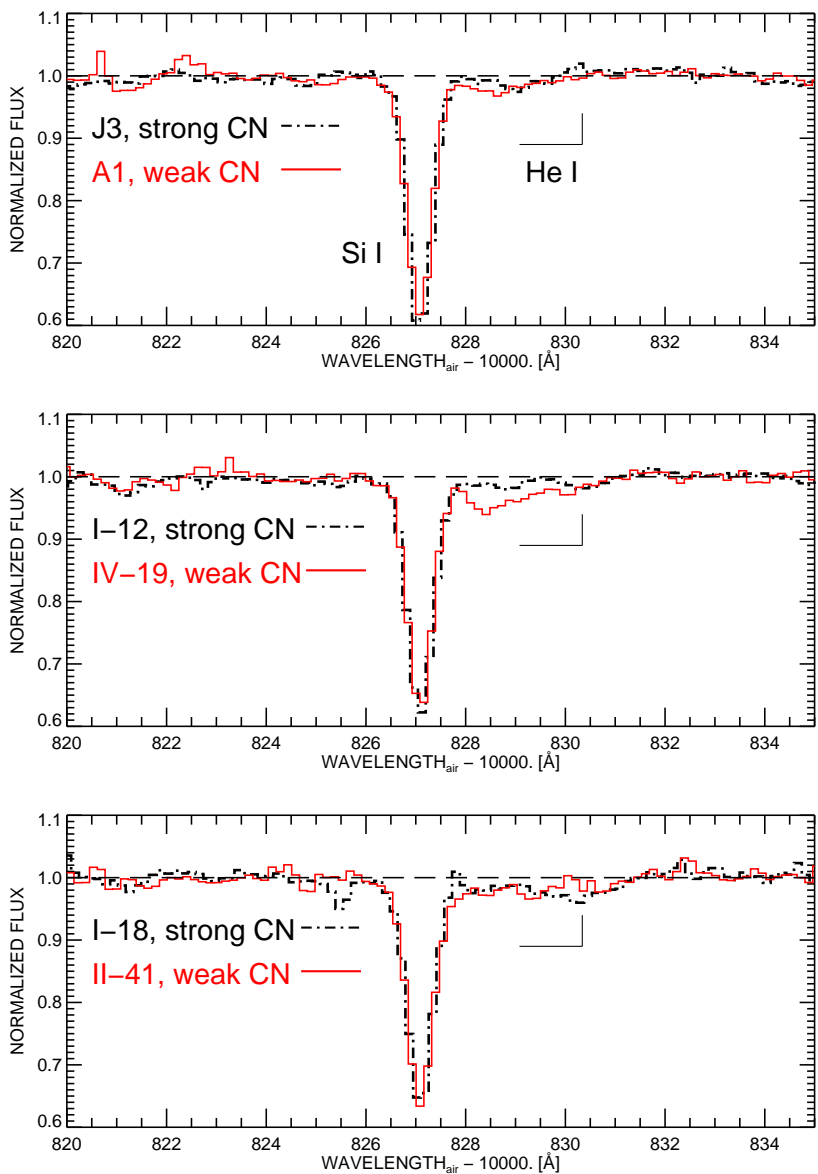}
\caption{Spectra covering the wavelength range $\lambda\lambda$10820-10835
\AA\ for three pairs of giants in M13. The horizontal axis plots the rest
frame wavelength minus 10000 \AA. The stars in each pair have similar
absolute visual magnitudes and $B-V$ colors but differ in $\lambda$3883 CN
band strength. In each panel the spectrum of the CN-strongest giant is shown
as a black (broken) line, while the spectrum of the CN-weaker giant is depicted
with a red (solid) line. The location of the $\lambda$10830 \ion{He}{1} 
feature is indicated by the horizontal line.}
\end{center}
\end{figure}
\clearpage

\begin{figure}
\figurenum{9}
\begin{center}
\includegraphics[scale=0.7]{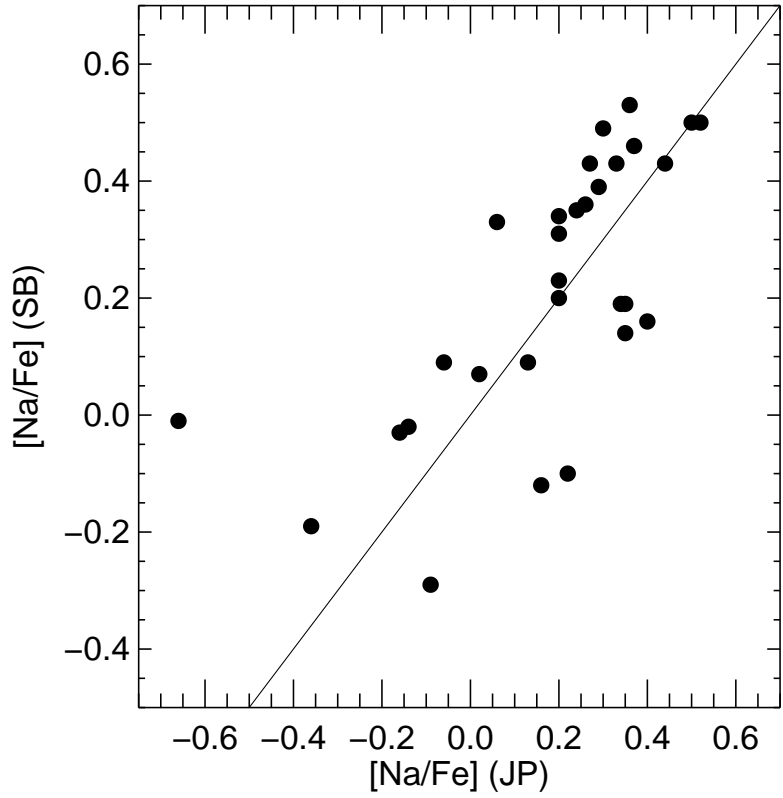}
\caption{The homogenized [Na/Fe] abundances from Smith \& Briley (2006) versus 
Johnson \& Pilachowski (2012) determinations. A straight line
depicts equality between the two abundance sets.}
\end{center}
\end{figure}
\clearpage

\begin{figure}
\figurenum{10}
\begin{center}
\includegraphics[scale=0.7]{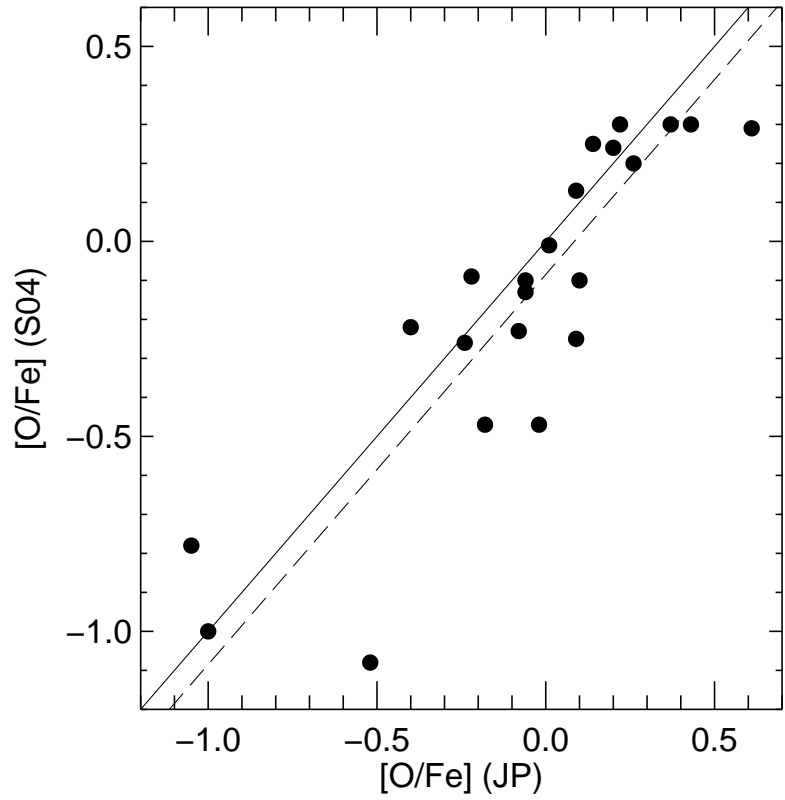}
\caption{Oxygen abundance data from Sneden et al. (2004) versus measurements  
from Johnson \& Pilachowski (2012). A straight line depicts equality 
between the two abundance sets, while the dashed line is offset by
0.085 dex.}
\end{center}
\end{figure}
\clearpage

\begin{figure}
\figurenum{11}
\begin{center}
\includegraphics[scale=0.7]{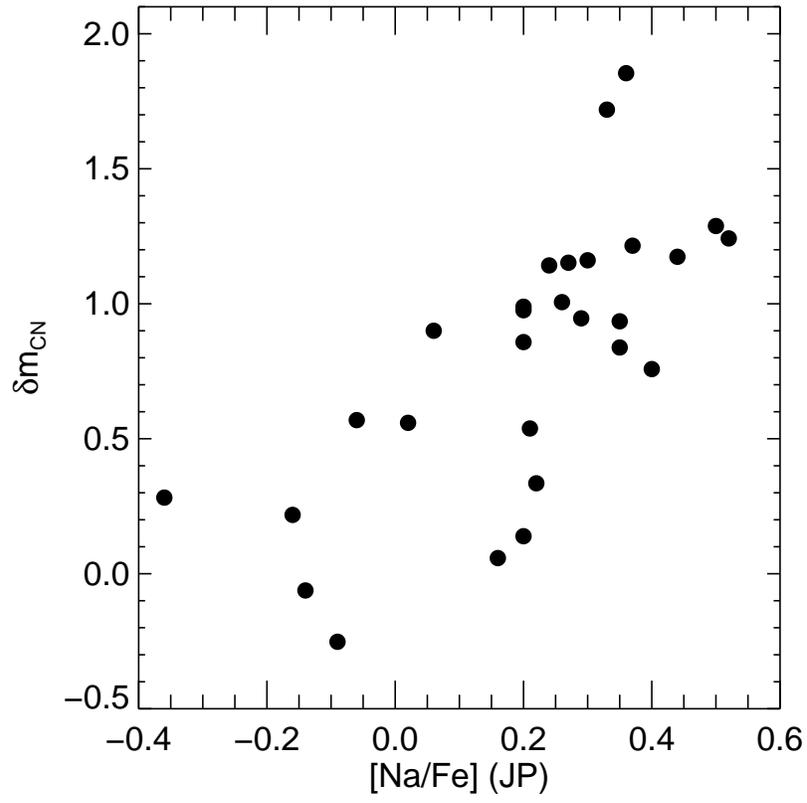}
\caption{The $\delta m_{CN}$ CN-band index versus [Na/Fe] measurements from 
Johnson \& Pilachowski (2012) for red giants in M13. There is a notable 
correlation.}
\end{center}
\end{figure}
\clearpage

\begin{figure}
\figurenum{12}
\begin{center}
\includegraphics[scale=0.7]{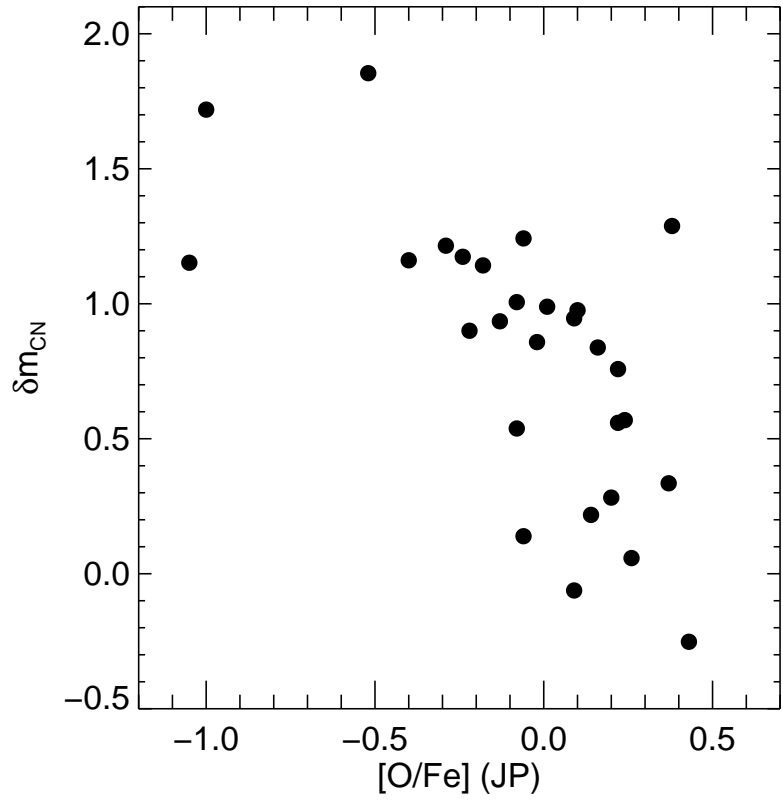}
\caption{The $\delta m_{CN}$ CN-band index versus [O/Fe] abundance from Johnson
\& Pilachowski (2012) for red giants in M13. There is a notable 
anticorrelation.}
\end{center}
\end{figure}
\clearpage

\newpage

\begin{deluxetable}{lccccccccc}
\tabletypesize{\small}
\tablecolumns{9}
\tablewidth{0pt}
\tablenum{1}
\tablecaption{Observational Data for Red Giants in Messier 13} 
\tablehead{
\colhead{Star}             &
\colhead{Exp}              &
\colhead{$V$}              &         
\colhead{$M_V$}            &
\colhead{$B-V$}            &      
\colhead{$m_{CN}$}         &    
\colhead{CN class}         &
\colhead{$EW_1$ (\ion{He}{1})}     &
\colhead{$EW_2$ (\ion{He}{1})}     &
\colhead{$EW$ (\ion{Si}{1})} \\
\colhead{}          &
\colhead{(s)}       &
\colhead{}          &         
\colhead{}          &
\colhead{}          &      
\colhead{}          &    
\colhead{}          &
\colhead{(m\AA) }    &
\colhead{(m\AA) }    &
\colhead{(m\AA) } 
}
\startdata
I-12    &  4800  &  13.54  &  $-0.79$  &  0.99  &  0.342  &  s  &  17.2  &  15.0   &  244.8 \\
I-18    &  6000  &  13.95  &  $-0.38$  &  0.92  &  0.333  &  s  &  72.0  &  80.8   &  236.8 \\ 
II-41   &  6000  &  13.95  &  $-0.38$  &  0.94  &  0.041  &  w  &  55.5  &  117.6  &  224.1 \\
IV-19   &  4800  &  13.68  &  $-0.65$  &  0.98  &  0.120  &  w  &  89.4  &  116.9  &  218.0 \\
A1      &  3600  &  13.39  &  $-0.94$  &  1.05  &  0.221  &  i  &  33.9  &  34.3   &  235.8 \\
J3      &  3600  &  13.31  &  $-1.02$  &  1.06  &  0.465  &  s  &  24.2  &  34.3   &  251.5 \\
X24     &  6000  &  13.75  &  $-0.58$  &  0.95  &  0.295  &  s  &  PCyg  &  PCyg   &  233.6 \\  
\enddata

\end{deluxetable}
\clearpage

\begin{deluxetable}{lrcrrrc}
\tabletypesize{\small}
\tablecolumns{7}
\tablewidth{0pt}
\tablenum{2}
\tablecaption{Abundance Data for Program Stars in Messier 13} 
\tablehead{
\colhead{Star}             &     
\colhead{$\delta m_{CN}$}  &    
\colhead{CN class}         &
\colhead{[O/Fe]}           &
\colhead{[Na/Fe]}          &
\colhead{[O/Fe]}           &
\colhead{[Na/Fe]}          \\
\colhead{}          &
\colhead{}          &
\colhead{}          &         
\colhead{SB\tablenotemark{a}}        &
\colhead{Sn04,SB\tablenotemark{b}}   &      
\colhead{JP\tablenotemark{c}}        &    
\colhead{JP}        \\
\colhead{1}         &
\colhead{2}         &
\colhead{3}         &         
\colhead{4}         &
\colhead{5}         &      
\colhead{6}         &    
\colhead{7}                  
}
\startdata
I-12    &     0.858  &   s  &  $-0.47$  &    0.20   &  $-0.02$  &   0.20  \\
I-18    &     0.981  &   s  &    ....   &    0.41   &    ....   &   ....  \\ 
II-41   &  $-0.142$  &   w  &    0.25   &  $-0.16$  &    ....   &   ....  \\
IV-19   &     0.058  &   w  &    0.20   &  $-0.12$  &    0.26   &   0.16  \\
A1      &     0.335  &   i  &    0.30   &  $-0.10$  &    0.37   &   0.22  \\
J3      &     1.242  &   s  &  $-0.10$  &    0.50   &  $-0.06$  &   0.52  \\
X24     &     0.758  &   s  &    ....   &    0.16   &    0.22   &   0.40  \\
\enddata

\tablenotetext{a}{Smith \& Briley (2006).}
\tablenotetext{b}{Abundances from Sneden et al. 2004 as tabulated in Smith \& 
Briley (2006).}
\tablenotetext{c}{Johnson \& Pilachowski (2012).}

\end{deluxetable}
\clearpage

\end{document}